\documentclass[aps,pra,twocolumn, superscriptaddress]{revtex4}
\usepackage{graphicx}
\usepackage{amsmath}
\begin{document}

\title{Amplitude to phase conversion of InGaAs pin photo-diodes for femtosecond lasers microwave signal generation}


\author{W. Zhang}
\affiliation{LNE-SYRTE, Observatoire de Paris, CNRS, UPMC, 75014 Paris, France}
\author{T. Li}
\affiliation{LNE-SYRTE, Observatoire de Paris, CNRS, UPMC, 75014 Paris, France}
\affiliation{Key Laboratory of Quantum Optics, Shanghai Institute of Optics and Fine Mechanics, Chinese Academy of Sciences, Shanghai 201800, PRC}
\author{M. Lours}
\affiliation{LNE-SYRTE, Observatoire de Paris, CNRS, UPMC, 75014 Paris, France}
\author{S. Seidelin}
\affiliation{Institut N\'eel, CNRS et Universit\'e Joseph Fourier, BP 166, F-38042 Grenoble Cedex 9, France}
\author{G. Santarelli}
\affiliation{LNE-SYRTE, Observatoire de Paris, CNRS, UPMC, 75014 Paris, France}
\author{Y. Le~Coq}
\email{yann.lecoq@obspm.fr}
\affiliation{LNE-SYRTE, Observatoire de Paris, CNRS, UPMC, 75014 Paris, France}



\begin{abstract}
When a photo-diode is illuminated by a pulse train from a femtosecond laser, it generates microwaves components at the harmonics of the repetition rate within its bandwidth. The phase of these components (relative to the optical pulse train) is known to be dependent on the optical energy per pulse. We present an experimental study of this dependence in InGaAs pin photo-diodes illuminated with ultra-short pulses generated by an Erbium-doped fiber based femtosecond laser. The energy to phase dependence is measured over a large range of impinging pulse energies near and above saturation for two typical detectors, commonly used in optical frequency metrology with femtosecond laser based optical frequency combs. When scanning the optical pulse energy, the coefficient which relates phase variations to energy variations is found to alternate between positive and negative values, with many (for high harmonics of the repetition rate) vanishing points. By operating the system near one of these vanishing points, the typical amplitude noise level of commercial-core fiber-based femtosecond lasers is sufficiently low to generate state-of-the-art ultra-low phase noise microwave signals, virtually immune to amplitude to phase conversion related noise.  
\end{abstract}

\maketitle


\section{Introduction}

Ultra-stable microwave signals are of great interest in a variety of applications such as radar, telecommunication and deep space navigation systems, timing distribution and synchronization \cite{Kim_Nature_08, Kim_OL_07}, ultra-high resolution Very Long-Baseline Interferometry, and development of local oscillators for accurate fountain atomic frequency standards \cite{Weyers_PRA_09, Millo_APL_09}. 

Femtosecond lasers have revolutionized the field of time and frequency metrology by providing a phase coherent link between optical and microwave frequencies \cite{Holzwarth_PRL_00, Ramond_OL_02}. By transferring the unrivaled spectral purity of modern ultra-stable continuous wave (cw) lasers \cite{Jiang_NP_11, Ludlow_OL_07, Webster_PRA_08, Millo_PRA_09, Leibrandt_OE_11} to the microwave domain {\it via} such a device \cite{Millo_OL_09, Zhang_IEEEUFFC_11}, one can generate extremely low phase noise reference signals \cite{Fortier_ARXIV_11} with numerous technological applications. The conversion from the optical to the microwave domain is based on the synchronization of the pulse repetition rate with the optical frequency of the cw laser \cite{Zhang_APL_10}. The subsequent detection of the optical pulse train, typically using a fast photo-diode, provides access to the microwave signal, which is a frequency-divided replica of the cw-laser reference. This process is, however, accompanied by excess phase noise which limits the residual timing stability of the microwave frequency generation \cite{McFerran_EL_05, Bartels_OL_05, Ivanov_UFFC_07}. One of the main causes for this excess phase noise is the amplitude to phase conversion in the fast photo-diode, combined with the unavoidable intensity noise of the femtosecond laser. 

In this paper, we examine the behavior of different photo-diodes commonly used for microwave generation, when illuminated by femtosecond pulses with repetition rate $f_{\rm rep} $\cite{Taylor_IEEEPJ_11}. Our goal is to measure and interpret the energy per pulse to microwave phase conversion (EPC) coefficient, defined as:
\begin{equation}
	{\rm EPC_{coef}} = \frac{d\phi(\omega,E)}{E.dE},
\end{equation}
where $\phi(\omega,E)$ is the spectral phase (relative to the optical pulse) of the photo-generated microwave component at angular frequency $\omega=2\pi.n.f_{\rm rep}$ ($n$ being an integer). Note that, from our definition of ${\rm EPC_{coef}}$, this quantity is expressed in [rad] per relative energy change which allows a mathematically simple transfer of the relative intensity noise of a femtosecond laser to the EPC-induced microwave phase noise. 

We begin by showing the measured time response of the photo-detector for a series of impinging optical energies per pulse. A simplistic model is presented, which allows one to understand the related ${\rm EPC_{coef}}$ behavior and gain insight into the best operating condition in the context of ultra-low phase noise microwave generation with optical frequency combs. We proceed by presenting some direct measurements of the amplitude to phase conversion factors of different photo-diodes under different illuminating conditions, which agrees qualitatively well with the predictions. We conclude by showing how these results can be applied to realize microwave generation with negligible EPC-induced noise over a large part of the Fourier spectrum. This paves the way to generating microwave signals with a phase noise level close the state-of-the-art of any competing technology, by using a fiber-based optical frequency comb phase locked to an ultra-stable cw laser. 


\section{Time response}

When applying a laser pulse whose duration is much shorter than 1\,ps to a pin photo-diode whose bandwidth is a few tens of GHz, one can consider (and we a have experimentally verified) that the time response does not depend significantly on the exact pulse shape and duration, as the bandwidth of the detector is by far the limiting factor. In addition, as long as the time response of the photo-diode is faster than $1/f_{\rm rep}$, one can consider that the photo-detector has enough time to get back to equilibrium (dark current) in between the pulses. We have applied short (fs regime) pulses at $f_{\rm rep}=250$\,MHz repetition rate from a Er-doped commercial fiber laser (MenloSystems GmbH M-comb) on 3 different photo-diodes: two DSC40S (14 GHz bandwidth), and one HLPD (High power handling capability, 14\,GHz bandwidth) from Discovery Semiconductor inc. We have recorded the time response at various optical powers (\emph{i.e.} various energies per pulse) using a 40\,GHz bandwidth sampling oscilloscope. Typical results obtained for the HLPD are presented in fig. 1. (top). At low energy per pulse, the responses of the photo-diodes are linear and the durations of the electronic pulses are limited by the bandwidth of the detector. When increasing the energy per pulse, the peak voltage increases linearly and the pulse distortion is, at first, very minute. When reaching saturation energies and above, the peak voltage saturates, and the electronic pulse takes a characteristic asymmetric shape where the fast rise is followed by a slow decrease down to equilibrium (considered here to be at 0\,V).

\begin{figure}
	\centering
	\includegraphics[width=8.7cm]{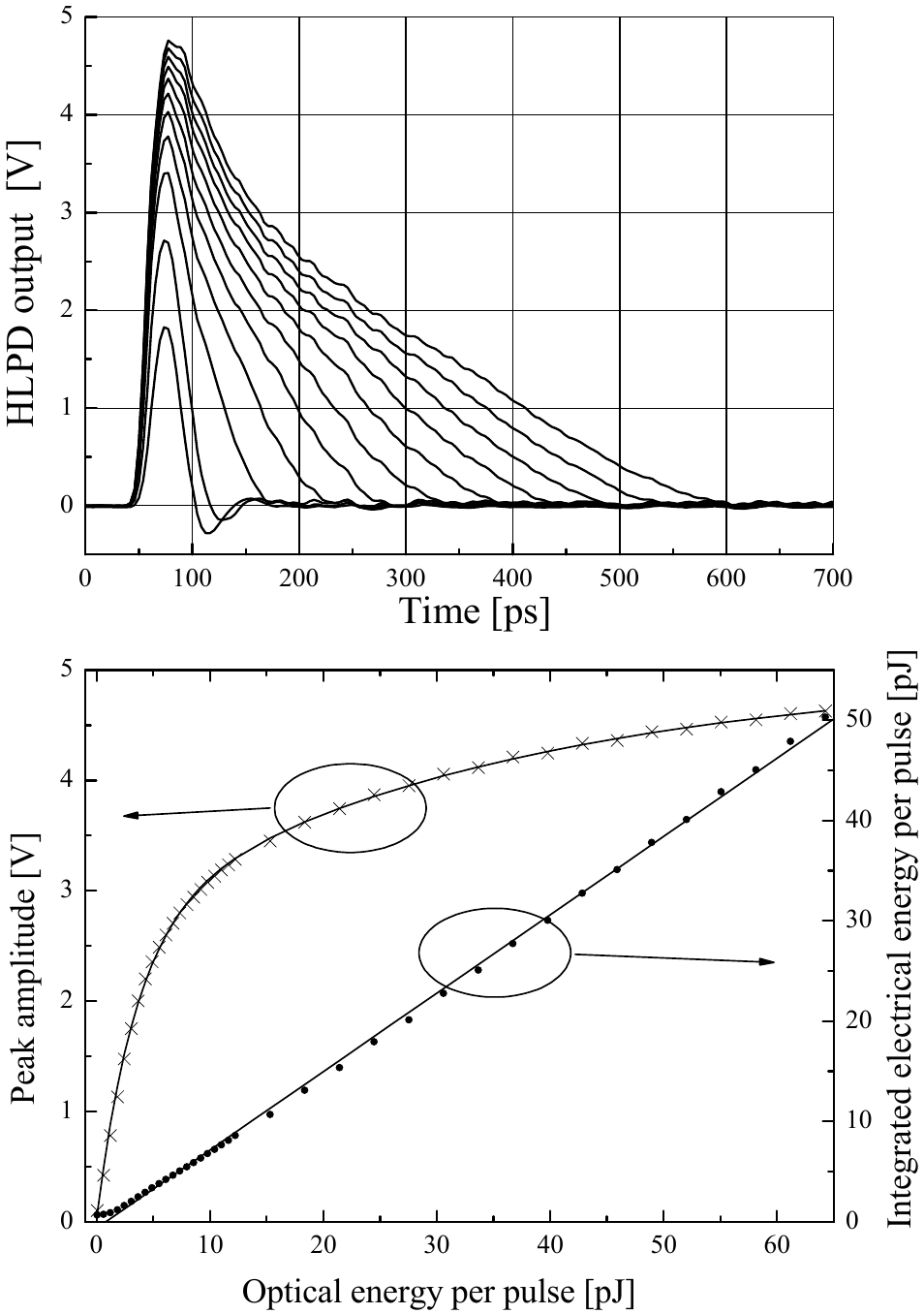}
	\caption{Time response of photo-diode HLPD. Top: typical time record for various optical energies per pulse, up to 64 pJ. Bottom left y-scale: peak voltage vs optical energy per pulse (the corresponding solid line is a double exponential fit), bottom right y-scale: integrated electrical energy vs optical energy per pulse (the solid line is a linear fit, exhibiting 78 percent quantum efficiency).}
	\label{fig:TimeSeries}
\end{figure}

The involved phenomenon is well known and related to space charge screening effects in the intrinsic part of the junction \cite{Kuhl_JLT_92, Williams_IEEEPTL_94, Dentan_JLT_90}. In short, the optical pulse is absorbed in the intrinsic part of the p-i-n junction where it is converted into a large number of electron-hole pairs (with a quantum efficiency $\eta$, around 0.8 for InGaAs photo-detector at 1.5 $\mu$m wavelength). Under the effect of the polarization bias field, the electrons and holes separate, and eventually recombine {\it via} the electronic circuit load. This gives rise to a time dependent photo-current which is measured on a 50\,ohms load. However, above saturation, the electric field induced by the carriers themselves is non-negligible, opposes the externally applied bias field, and induces a decrease of the mobility of the carrier. This phenomenon explains the slow decrease of the signal at high optical energies per pulse.

Fig \ref{fig:TimeSeries} (bottom, left y-axis) shows the peak voltage {\it vs}. optical energy per pulse, which exhibits a characteristic saturation-shape which is empirically well fitted by a double exponential. Fig \ref{fig:TimeSeries} (bottom, right y-axis) shows the integrated electrical energy per pulse {\it vs.} optical energy per pulse, allowing one to measure the quantum efficiency, which is observed to be constant over the full data range.


\section{Simple model}

From our measurements and previous literature on the subject \cite{Ivanov_IEEEUFFC_05, Joshi_PTL_09}, we can infer a simplistic model for the time response of pin photo-diodes under illumination by femtosecond pulses above saturation. The model proves sufficient to gain insight into the impact of the asymmetric time response on the EPC. For a given optical energy per pulse $E$, the time response photo-current $i(t)$ can be approximated by an asymmetric triangular response model (thereafter abbreviated ATRM):
\begin{equation}
i(t) = \left\{
    \begin{array}{ll}
        i_{\rm max} \cdot (1-t/\tau) & \mbox{for } 0<t<\tau, \\
        0 & \mbox{elsewhere},
    \end{array}
\right.
\end{equation}
where $i_{\rm max}$ and $\tau$ are monotonous increasing functions of $E$. In practice, we let (empirically) $i_{\rm max}$ follow a double exponential growth with $E$:
\begin{equation}
	i_{\rm max} = i_0 \cdot \left( 1 - A_1 \cdot e^{-E/E1} - A_2 \cdot e^{-E/E_2} \right), 
\end{equation}
with $A_1+A_2=1$. The dependence of $\tau$ with $E$ is found by energy conservation: the total energy of the electric pulse is equal to the {\it absorbed} optical energy $\eta E$. Therefore, in a $R=50\,\Omega$ load resistance, $\eta E = R i_{\rm max}^2(E) \tau(E) / 3$, which gives:
\begin{equation}
 	\tau = 3 \eta E / R i_0^2(1-A_1 e^{-E/E1}-A_2e^{-E/E2})^2.
\end{equation}
For a given photo-diode, the parameters $i_0$, $A_1$, $A_2$, $E_1$ and $E_2$ can be determined experimentally by a double-exponential fit of the peak photo-current, and the parameter $\eta$ by a linear fit of the integrated energy of the electrical pulse. As example, both fits are represented in the bottom plot of fig. \ref{fig:TimeSeries} for HLPD. 

Very generally, in the frequency domain, the spectral phase associated with the angular frequency $\omega$ is 
\begin{equation}
	\phi(\omega) = \left. \int_{-\infty}^{+\infty} i(t) \sin(\omega t) \,dt \middle/ \int_{-\infty}^{+\infty} i(t) \cos(\omega t) \,dt \right. ,
\end{equation}
which leads, within the ATRM to:
\begin{equation}
	\phi(\omega) = \arctan \left[\frac{\omega\tau-\sin(\omega\tau)}{1-\cos(\omega\tau)} \right],
\end{equation}
and therefore:
\begin{equation}
	{\rm EPC_{coef}} =
	\frac{\omega\tau\sin(\omega\tau)+2\cos(\omega\tau)-2}{2\omega\tau\sin(\omega\tau)+2\cos(\omega\tau)-\omega^2\tau^2-2}
	\times \frac{\omega\,d\tau}{dE/E}.
\end{equation}

For a given angular frequency $\omega$, this coefficient oscillates between positive and negative values when increasing $E$ and, interestingly, vanishes for pulse energies such that $\omega\tau/2$ is a solution of the equation $\tan(x)=x$. Consequently, given an angular frequency $\omega$ and a range of physically accessible energies per pulse (from 0 to $E_{\rm max}$), a photo-diode will exhibit a number of ${\rm EPC_{coef}}$ vanishing energies which increases with both $\omega$ and $E_{\rm max}$. More precisely, from the ATRM, at an angular frequency $\omega$, the number of energy values between 0 and $E_{\rm max}$ for which ${\rm EPC_{coef}}$ vanishes is roughly $N_0(E_{\rm max}) \simeq \omega\tau(E_{\rm max})/\pi-3/2$ (not counting the trivial case of zero energy per pulse). These vanishing points are particularly interesting in the context of low phase noise microwave generation: with a pulsed laser prone to amplitude noise, they nevertheless allow microwave signal generation without EPC-induced phase noise degradation during photo-detection. Physically, these vanishing points appear only above saturation, when the time response of the photo-diode starts to exhibit the strongly asymmetric shape which is described simplistically by the ATRM.


\section{Experimental results}

Beyond the simplistic ATRM description, from the time series of fig. \ref{fig:TimeSeries}.a, one could in principle numerically compute ${\rm EPC_{coef}}$ for various microwave angular frequencies $\omega$. The timing resolution of the sampling oscilloscope and the necessity to register a high density of time responses to obtain a large enough data sample renders this method difficult and cumbersome to implement. We have however verified qualitatively by such a technique that ${\rm EPC_{coef}}$ alternates between positive and negative values when varying $E$, as well as the roughly linear increase of $N(E_{\rm max},\omega)$ for increasing $\omega$.

\begin{figure}[htb]
	\centering
	\includegraphics[width=8.7cm]{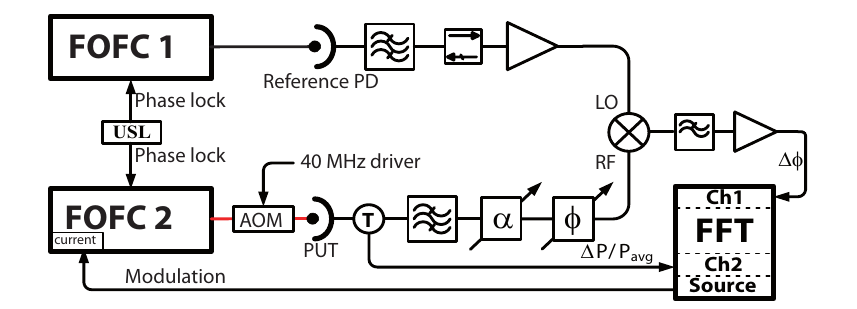}
	\caption{Setup schematics for the direct measurement of EPC coefficients. FOFC: Erbium doped fiber-based optical frequency comb; USL: ultra-stable laser; AOM: Acousto-optic modulator; FFT: Fast Fourier transform analyzer; Reference PD: reference signal photo-diode; PUT: photo-diode under test; LO: local oscillator input of the mixer; RF: radio-frequency input of the mixer.}
	\label{fig:AM2PMSetup}
\end{figure}

An alternative way to know ${\rm EPC_{coef}}$ for a given microwave angular frequency $\omega$ is to measure it directly with the setup depicted in fig. \ref{fig:AM2PMSetup}. Briefly, two quasi-identical commercial-core erbium-doped fiber-based optical frequency combs are phase locked to the same ultra-stable cw reference signal \cite{Zhang_IEEEUFFC_11} by acting on a piezo-actuator controlling the length of the oscillator's cavities. For one of the combs, the pulse train output (250\,MHz repetition rate) passes through an acousto-optic modulator, used as a variable attenuator which allows hysteresis-free very fine tuning of the power, and hence of the energy per pulse (approx. 10\,fJ level resolution). After the AOM, the maximum available power is 30\,mW, which corresponds to 120\,pJ/pulse. After photo-detection by the fast InGaAs pin photo-diode under test (PUT), the signal is attenuated by a broadband resistive attenuator (-30\,dB), and sent to a narrow bandpass filter which isolate the harmonic of the repetition rate at angular frequency $\omega$. The attenuation ensures broadband impedance matching and isolation of the PUT from the components reflected by the narrow filter. After filtering, the signal (angular frequency $\omega$, power dependent on the configuration but always below -40\,dBm) is sent to the RF port of a double-balanced microwave mixer. The second comb's pulse train, used to generate a reference microwave signal, is similarly photo-detected (although without variable optical attenuation), bandpass filtered and, this time, amplified to saturate the LO port of the mixer. The attenuation of the first microwave signal to below -40\,dBm and saturation of the LO port ensures a linear operation of the mixer, where the amplitude-phase conversion of the mixer itself is negligible. The two combs, although phase locked to the same cw reference are operating at slightly different repetition rates, such that the two generated microwave signals exhibit a beatnote (at the IF port of the mixer) typically of a few tens of kHz. A Fast Fourier Transform analyzer (FFT, Agilent 89410A) working in analog demodulation mode is tracking this beatnote and analyzing both its amplitude phase fluctuations (``FFT channel 1'' in fig. 2). This approach does not require phase calibration. Moreover, the instrument has a negligible amplitude to phase coupling.

To generate a pure amplitude modulation on the PUT, we apply a small sinewave modulation to the current driving one of the pump diode lasers of the corresponding femtosecond laser. The frequency of this modulation is chosen slow enough (30\,Hz) to be well within the bandwidth of the comb servo loop which phase locks it to the optical cw reference. As a consequence, the resulting phase modulation is negligible, as the main phase lock loop is constantly compensating for the modulation. With this method, the optical signal impinging on the PUT exhibits a quasi-pure amplitude modulation at a low frequency, which turns into a microwave phase modulation after photo-detection, due to EPC.

To calibrate the optical amplitude fluctuation, we insert after the PUT a bias T whose near-DC output is sent to the FFT analyzer (``FFT channel 2'' in fig. \ref{fig:AM2PMSetup}). The FFT analyzer exhibits a peak at 30Hz Fourier frequency (baseband demodulation operating mode) whose amplitude corresponds to the optical amplitude modulation {\it via} calibration of the photo-diode. In addition, a measurement of the average value by a voltmeter in parallel gives the average energy per pulse (as the repetition rate is known to be 250\,MHz).

\begin{figure}
	\centering
	\includegraphics[width=8.7cm]{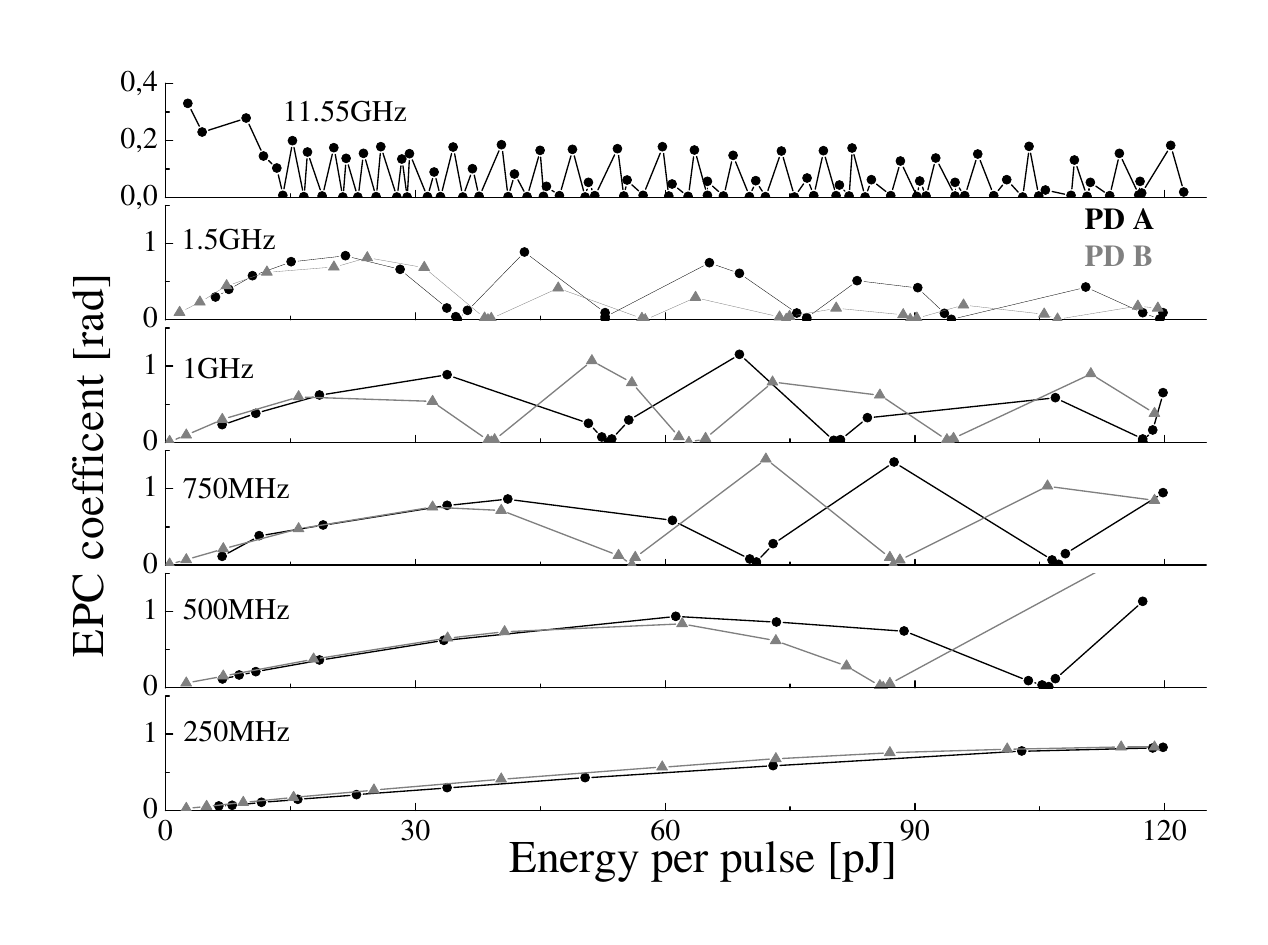}
	\caption{Amplitude to phase conversion coefficients obtained for two DSC40S photo-diodes from Discovery Semiconductor Inc at various carrier frequencies. Only the absolute value of the coefficient is plotted. The conversion factor was verified independently to take alternatively positive and negative coefficients on each sides of the vanishing points. For clarity only one photo-diode results is shown for 11.55\,GHz carrier frequency. The number of zeros within the accessible range of energies per pulse (respectively 0, 1, 2, 3, 5 and 45 ) agrees well with the ATRM predictions (see text).}
	\label{fig:AMPMDSC40S}
\end{figure}

\begin{figure}
	\centering
	\includegraphics[width=8.7cm]{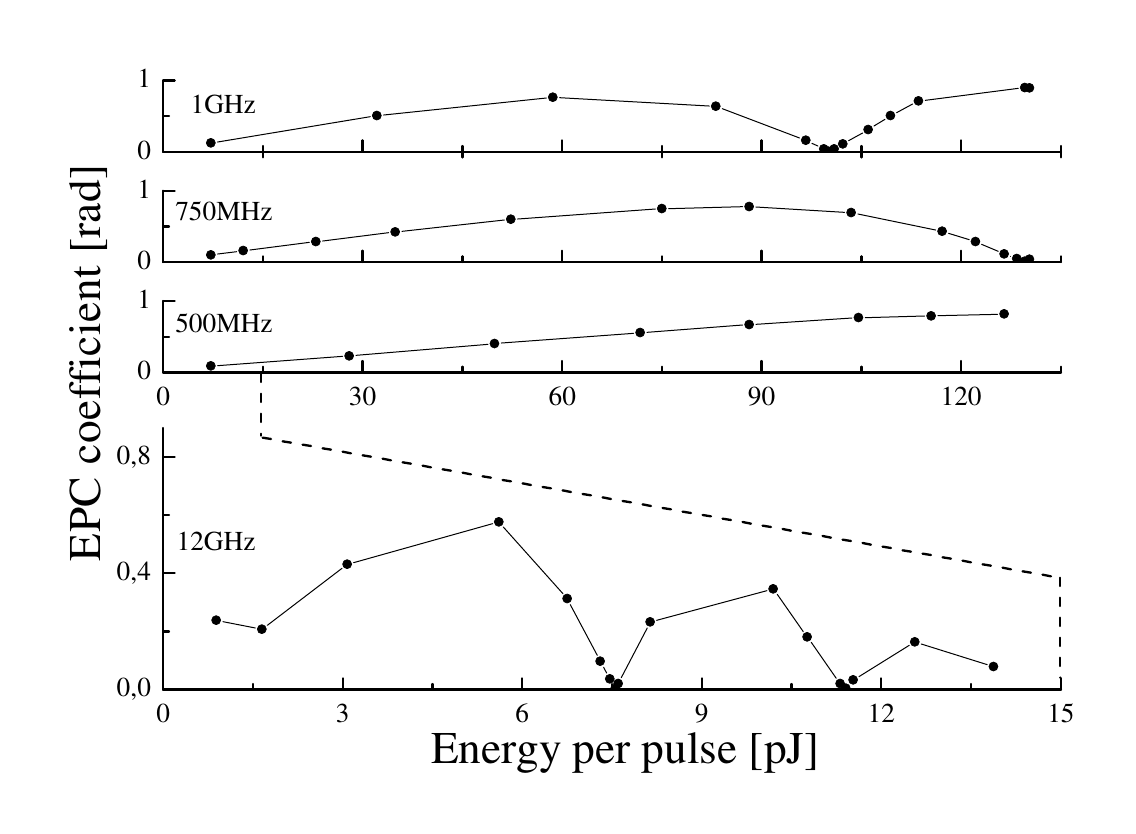}
	\caption{Amplitude to phase conversion coefficients obtained for the HLPD photo-diode at various carrier frequencies. Only the absolute value of the coefficient is plotted. the temperature is servo-ed at 19\,$^\circ$C and the bias voltage at 9\,V.}
	\label{fig:AMPMHLPD}
\end{figure}

Example of such measurements, realized for two DSC40S (nominally identical) and one HLPD photo-diodes from Discovery Semiconductor Inc. are presented in fig. \ref{fig:AMPMDSC40S} and \ref{fig:AMPMHLPD}. The measurements obtained using this method confirms the ATRM prediction of multiple zeros of ${\rm EPC_{coef}}$ with a roughly linear increase of $N(E_{\rm max},\omega)$ with $\omega$, where $N(E_{\rm max},\omega)$ is the number of ${\rm EPC_{coef}}$ zeros within the explored energy range $[0;E_{\rm max}]$.  More quantitatively, from independent measurements of the time response similar to that of figure \ref{fig:TimeSeries}, we deduce, in the ATRM framework for  $E_{\rm max}=$120\,pJ, for the two DSC40S  as well as for the HLPD, a number of EPC vanishing points $N_0(E_{\rm max})$ which agrees with the actual measurement.

We have observed experimentally that the exact values of the energy per pulse which cancel out the EPC are very dependent of the exact PUT, even within the same model number (see for example fig. \ref{fig:AMPMDSC40S} for two nominally identical DSC40S). Each PUT must therefore be independently characterized to know the exact location of such zeros.

In order to use these EPC vanishing points for EPC-free microwave generation, it is important to verify that they are sufficiently stable over time and operating conditions. We have realized a series of measurements for that purpose where we focus on the first EPC vanishing point of the HLPD photo-diode at 12 GHz carrier (which is near 7.5\,pJ per pulse in the device we have tested at 19\,$^\circ$C temperature and 9\,V bias, see fig. \ref{fig:AMPMHLPD}). When varying the temperature of the photo-diode from 14\,$^\circ$C to 26\,$^\circ$C, the EPC vanishing point moves by less than 10\,fJ/\,$^\circ$C. The exact optical pulse shape and spectrum can, in practice, be slightly modified for different mode-lock states of the femtosecond laser. We tried three different mode-lock states of the laser, which did not produce any significant displacement of the EPC vanishing point. Furthermore, using the natural dispersion of optical fibers, the duration of the pulse can be easily increased. We tried adding 3\,m, 6\,m and 12\,m of fiber length before the photo-diode, which did not move significantly the first EPC vanishing point. It is worth noticing that the second vanishing point did however exhibit a small but significant shift  for 12m of added fiber length. Using a fibered polarization controller, we have also varied the input polarization of the light without any discernible effect on the first EPC vanishing point position. We also changed the repetition rate of the fiber comb (by up to 0.1 percent) without discernible change. In addition, we have tried to realize the measurement at different modulation frequencies (10\,Hz, 30\,Hz, 300\, Hz and 1\,kHz) without any effect, either on the amplitude or the dependence with $E$ of ${\rm EPC_{coef}}$. This last result directly contradicts reference \cite{Wu_IEEEPTL_11}. The discrepancy is probably due to the free-running laser method used in this reference, which we believe to be inadequate. The strongest dependence of the first zero position for the studied HLPD is found to be with the bias voltage (which is normally a well controlled parameter) for which we measured a 0.7\,pJ/V shift. Repetitions of the same measurements after one week and after 3 months of time have produced no significant difference in the absolute position of the first EPC vanishing point or in its sensitivity to external conditions.

The last parameter to verify is the long term amplitude drift of the femtosecond laser itself. We have measured the amplitude of our free-running femtosecond laser to drift by less than 0.1 percent after 3 days. Combined with the measured slope of ${\rm EPC_{coef}}$ {\it vs.} $E$ for HLPD near its first EPC vanishing point (approx. 0.5\,rad/pJ) this implies a free running drift of ${\rm EPC_{coef}}$ from an initially nulled value to less than 0.003\,rad after 3 days. Overall, it seems therefore reasonable to guarantee, with an all passive system initially tuned to operate in the vicinity of an EPC vanishing point, an EPC factor which remains lower than 0.03\,rad per relative change over long periods of time (allowing up to 1 percent amplitude fluctuations over time). Such low factor would guaranty the EPC induced phase noise to be 30\,dB lower than the relative intensity noise (RIN) of the femtosecond laser. A more complicated system which would actively maintain the energy per pulse very close to a ${\rm EPC_{coef}}$ vanishing point may produce even higher rejection of laser amplitude noise, for very long times.


\section{Noise floor limits}

\begin{figure}
	\centering
	\includegraphics[width=8.7cm]{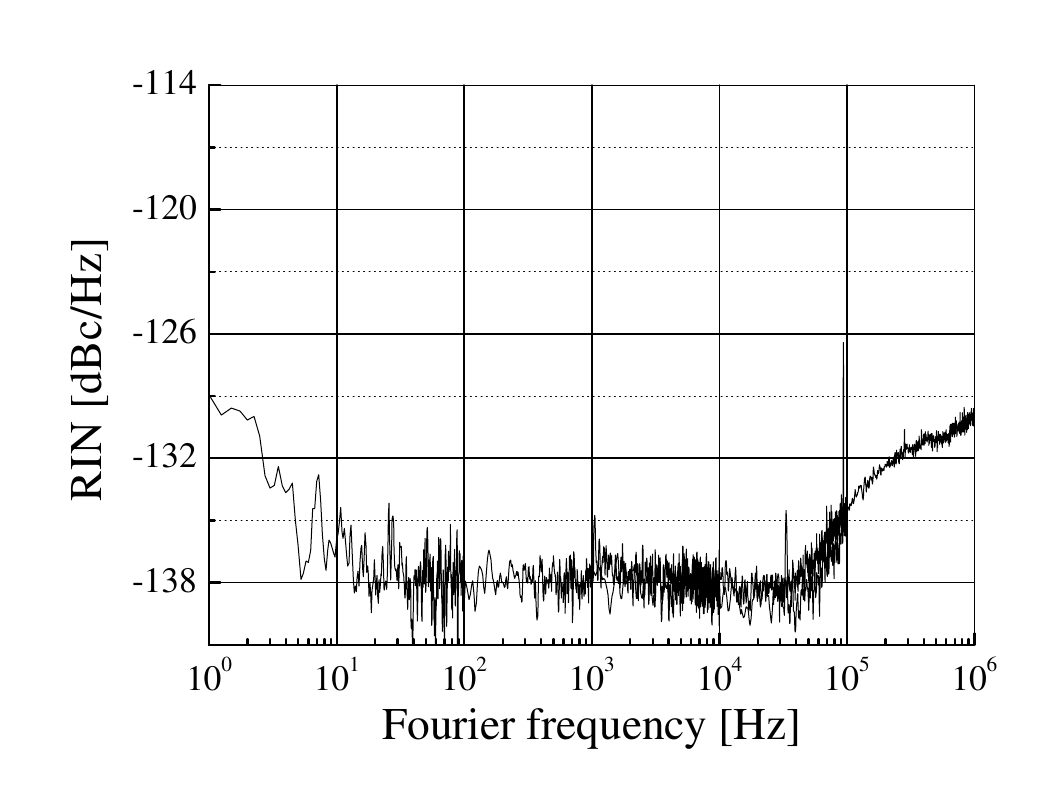}
	\caption{Typical RIN of the laser with home-made pump diodes power supplies. The small but significant increase near 1\,MHz Fourier frequency remains unexplained, has been observed to be mode-lock dependent and needs further investigation. The residual current noise of the fiber comb's pump diode lasers power supplies (-154\,dBc/Hz over the full range of Fourier frequencies) is not the leading cause of the measured laser noise, as proven by independent transfer function analysis.} 
	\label{fig:RIN}
\end{figure}

\begin{figure}
	\centering
	\includegraphics[width=8.7cm]{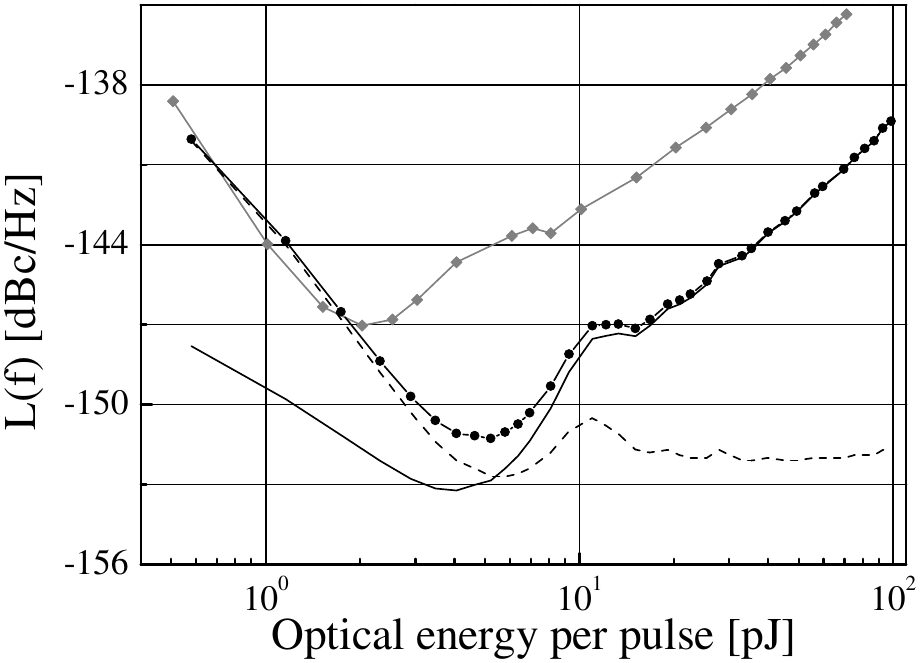}
	\caption{White noise limit inferred from measurements for two different photo-diodes at a carrier frequency of 12\,GHz, for various optical energies per pulse at a 250\,MHz repetition rate. The Johnson-Nyquist noise is calculated from 300\,K temperature and the measured microwave power at 12\,GHz. The shot noise is calculated from the measured average photo-current and the measured microwave power at 12\,GHz. The quadratic sum of Johnson-Nyquist and shot noise is giving the total white noise floor level which is shown for the two PUTs. The gray diamonds are for one DSC40S; the black circles are for the HLPD. For the later, the components of the white noise floor due to Johnson-Nyquist noise (black dashed line) and shot noise (black continuous line) are also represented separately.}
	\label{fig:ElectronicNoise}
\end{figure}

The RIN we typically measure from our commercial femtosecond laser (although with home-made low noise pump diodes power supplies) is about or below -130\,dBc/Hz (see fig; \ref{fig:RIN}). The combination of such low RIN with close to EPC vanishing points operation of the fast InGaAs photo-diode implies that the EPC-induced phase noise degradation of the microwave extraction process is below -160\,dBc/Hz level for Fourier frequencies between 1\,Hz and 1\,MHz, a level at which the Johnson-Nyquist noise and shot noise are, by far, the main limiting factors (see fig \ref{fig:ElectronicNoise}). As a matter of fact, the limited amount of microwave power at a given harmonic of the repetition rate (typically about -30\,dBm) must be compared to a constant level of -177\,dBm of phase noise arising solely from Johnson-Nyquist noise at 300\,K temperature. Furthermore, the average photo-current produced by the PUT is also producing a shot noise whose amplitude becomes dominant at high optical powers. When considering both the EPC-induced noise, and the white noise (\emph{i.e.} the Johnson-Nyquist and shot noise together), a hypothetic optimum operating condition of a given PUT (in terms of optical energy per pulse) would be to find an EPC vanishing point in the vicinity of the minimum white noise floor represented in fig. \ref{fig:ElectronicNoise}. However, for microwave generation near 12\,GHz, comparing fig. \ref{fig:AMPMHLPD} and fig. \ref{fig:ElectronicNoise} shows that such ideal situation is not directly possible for the HLPD (the first EPC vanishing points arise at an optical energy per pulse significantly above the white noise floor minimum). We have observed the same situation for the two DSC40S photo-diodes we tested and believe it to be similar for other types of photo-diodes using the same technology. In fact, for white noise floor consideration, the best EPC vanishing point to be used is the one happening at lowest energy, which corresponds to the onset of saturation. This EPC vanishing point is, however, systematically slightly above the energy per pulse which produces the lowest white noise floor limit. This slightly degraded performance with respect to white noise is, however, a small cost to pay for the elimination of EPC-induced phase noise. The best noise floor we can obtain with our 250\,MHz repetition rate laser in EPC-free power tuning is therefore about $-150$\,dBc/Hz at 12\,GHz carrier, by using the HLPD which exhibits a noise floor significantly lower than the two DSC40S photo-diodes we tested. Further development is needed for lowering the noise floor limit beyond this point. 

\section{Conclusion}

Repetition rate multiplication by use of filtering Fabry-perot cavity \cite{Kirchner_OL_09, Steinmetz_APB_09} or other structures \cite{Preciato_OE_08, Magne_PHD_07} may provide a convenient way to improve on Johnson-Nyquist and shot noise limitation by increasing the microwave signal at a given carrier frequency. Intuitively, repetition rate multiplication techniques would, indeed, essentially scale down the white noise floor of figure \ref{fig:ElectronicNoise} by redistributing the optical energy from unused harmonics of the repetition rate to a smaller number of harmonics including the relevant carrier frequency. As long as the repetition rate multiplication technique does not distort the optical pulse to the point of making its duration comparable or larger than the bandwidth of the detector, the EPC coefficient will not be modified by those technique (as we expressed it in term of optical energy per pulse and therefore independent of the actual signal's repetition rate). It should be possible to find a repetition rate multiplication factor for which the white noise at the first EPC vanishing point is comparable or lower than the residual EPC-induced phase noise. However, repetition rate multiplication techniques come at the expense of a reduced energy per pulse which, in a practical case where the total optical power available from the source is necessarily limited, may become too low for reaching an EPC vanishing point regime. Furthermore, a higher repetition rate with a constant energy per pulse (corresponding ideally to the first EPC vanishing point for close-to-optimal white noise floor) implies a higher average optical power which may exceed the destruction threshold of the photo-diode. Under the assumption that a sufficiently high optical power is available from the laser source, the HLPD has a strong advantage over the DSC40S as its higher power handling capability allows both a lower white noise limit and a higher average optical power breakdown value. With this photo-diode, it seems reasonable to envision in the near future, with those combined techniques, a microwave extraction process limited at a level close to -160\,dBc/Hz at 12\,GHz carrier for high enough Fourier frequencies (above 10\,kHz if one assumes a low Fourier frequency behavior identical to \cite{Zhang_IEEEUFFC_11, Zhang_APL_10}). This would correspond to the very best of any competing technology.




\end{document}